\documentclass[prb,floatfix,superscriptaddress,showpacs,twocolumn]{revtex4}
\usepackage{bm}
\usepackage[colorlinks=true,linkcolor=blue]{hyperref}
\usepackage{amsmath}
\usepackage{amssymb}
\usepackage{amsthm}
\usepackage{amsfonts}
\usepackage{times}
\usepackage[caption=false]{subfig}
\usepackage{placeins}
\usepackage{latexsym}
\usepackage{graphicx}

\newcommand{\alao}{$a_{\rm LAO}$}
\newcommand{\asto}{$a_{\rm STO}$}
\newcommand{\adso}{$a_{\rm DSO}$}

\newcommand{\lao}{LaAlO$_3$}
\newcommand{\lno}{LaNiO$_3$}

\newcommand{\nith}{Ni$^{3+}$}
\newcommand{\dzt}{$d_{3z^2-r^2}$}
\newcommand{\dxtyt}{$d_{x^2-y^2}$}
\newcommand{\one}{(LaAlO$_3$)$_1$/(LaNiO$_3$)$_1$} 
\newcommand{\three}{(LaAlO$_3$)$_3$/(LaNiO$_3$)$_3$}
\newcommand{\muB}{$\mu_{\rm B}$}
\newcommand{\ef}{$E_{\rm F}$}

\usepackage{color}

\begin{document}

\title{Confinement induced metal-to-insulator transition in strained  LaNiO$_3$/LaAlO$_3$ superlattices}
\author{Ariadna Blanca-Romero}
\affiliation{Department of Earth and Environmental Sciences, Section Crystallography and Center of Nanoscience,
University of Munich, Theresienstr. 41, 80333 Munich, Germany}
\author{Rossitza Pentcheva}
\email{rossitzap@lmu.de}
\affiliation{Department of Earth and Environmental Sciences, Section Crystallography and Center of Nanoscience,
University of Munich, Theresienstr. 41, 80333 Munich, Germany}

\date{\today}
\pacs{73.20.-r,68.65.Cd,71.30.+h,74.78.Fk}

\begin{abstract}
Using density functional theory calculations including a Hubbard $U$ term we explore the effect of strain and confinement on the electronic ground state of superlattices containing the band insulator LaAlO$_3$ and the correlated metal LaNiO$_3$. Besides a  suppression of holes at the apical oxygen, a central feature is the asymmetric response to strain in single unit cell superlattices: For tensile strain a band gap opens due to charge disproportionation at the Ni sites with two distinct magnetic moments of 1.45$\mu_{\rm B}$ and 0.71$\mu_{\rm B}$. Under compressive stain, charge disproportionation is nearly quenched and the band gap collapses due to overlap of $d_{3z^2-r^2}$ bands through a semimetallic state.  This asymmetry in the electronic behavior is associated with the difference in octahedral distortions and rotations under tensile and compressive strain. The ligand hole density and the metallic state are quickly restored with increasing thickness of the (LaAlO$_3$)$_n$/(LaNiO$_3$)$_n$ superlattice  from $n=1$ to $n=3$.    

\end{abstract}

\maketitle
The ability to grow transition metal oxide heterostructures with atomic control of layer thickness opens an exciting realm to explore confinement effects and novel effectively two-dimensional electronic phases that are not accessible in the bulk compounds.
To this end, Chaloupka and Khaliullin~\cite{chaloupka08} recently proposed that a cuprate-like behavior may be stabilized in a La$_2$Ni$M$O$_6$ superlattice (SL) where  NiO$_2$ and
$M$O$_2$ planes alternate along the $c$-axis of the ABO$_3$ perovskite lattice and M denotes a trivalent
cation ($M$=Al, Ga, Ti).  LaNiO$_3$ is a perovskite compound with rhombohedral symmetry ($R\bar{3}c$) that remains metallic at all temperatures  due to degenerate $e_g$ orbitals at the Ni$^{3+}$-sites  ($t^6_{2g}e^1_g, S = 1/2$). In contrast to bulk LNO, all other  representatives of the rare earth nickelates undergo a metal-to-insulator transition (MIT)~\cite{torrance92} at low temperatures, driven by charge ordering (CO) rather than the Jahn-Teller effect.\cite{medarde97,mazin07} Chaloupka and Khaliullin~\cite{chaloupka08} proposed that the orbital degeneracy in LNO/LMO superlattices can be lifted and a selective \dzt\ or \dxtyt\ occupation at the \nith\ sites  achieved by applying strain induced by the substrate. Subsequent density functional theory (DFT) calculations in the local density approximation (LDA) found a two sheet Fermi surface (FS) with strongly hybridized \dzt\ and \dxtyt\ orbitals.\cite{hansmann09,hansmann10}  A single-sheet FS could only be obtained by including correlation effects within the dynamical mean field approximation (DMFT) and/or by using Sc as counterion to Ni instead of Al.\cite{andersen} Han \emph{et al.}~\cite{millis} showed that the choice of the counterion (Al, Ga, B, In) rather than strain can be used to control chemically the orbital polarization.

The DFT  studies mentioned above have considered exclusively superlattices in a tetragonal setup. However, both \lno\ (LNO) and \lao\ (LAO) are  rhombohedral in the bulk (space group $R\bar{3}c$). Here we explore the effect of confinement and strain on the octahedral distortions. DFT calculations with an on-site Coulomb  repulsion term show that the lattice distortions have a significant influence on the bandwidth and thereby on the electronic ground state of the system. In particular, a charge disproportionated (CD) insulating state is obtained in LAO$_1$/LNO$_1$ SLs under tensile strain. In a tetragonal cell such a state is reported only at higher $U$ values.~\cite{andersen}  The tendency toward CD is vastly suppressed under compressive strain.

A metal-to-insulator transition as a function of LNO thickness has been observed experimentally for \lno/\lao~\cite{liu2011,keimer} and \lno/SrMnO$_3$~\cite{may2009} superlattices as well as in thin LNO films grown on an STO(001) substrate.~\cite{triscone2011,stemmer2010} The Ruddlesden-Popper compound La$_4$Ni$_3$O$_8$~\cite{pardo2010,poltavets2010} is another example for an insulating state due to confinement. Here, we explore the effect of quantum confinement  by varying the period of the  (LAO)$_n$/(LNO)$_n$ superlattice from  $n=1$ to $n=3$. The results show  a quick convergence toward bulk metallic behavior within the LNO part of the superlattice with increasing thickness.
\section{Calculational Details}
To gain insight into the electronic behavior of (LAO)$_n$/(LNO)$_n$ ($n=1,3$) superlattices, we performed density functional theory calculations with the full potential linearized augmented plane wave method as implemented in the WIEN2k code.~\cite{wien} An on-site Coulomb repulsion term is considered beyond the generalized gradient approximation~\cite{pbe96} of the exchange correlation potential within the LDA/GGA+$U$ approach~\cite{anisimov93} using $U=4$~eV and $J=0.7$~eV for Ni and $U=7$ and $J=0$~eV for La. The muffin tin (MT) radii are $1.80~\textrm{bohrs}$ (Ni and Al), $2.30~\textrm{bohrs}$ (La) and $1.60~\textrm{bohrs}$ (oxygen). Inside the muffin tins, wave functions are expanded in spherical harmonics up to $l_{\rm max}^{\rm wf}=10$ and nonspherical contributions to the electron density and potential are considered up to $l_{\rm max}^{\rm pot}=6$. The energy cutoff for the plane wave representation in the interstitial is $E_{\rm max}^{\rm wf}=25$~Ry for the wave functions and $E_{\rm max}^{\rm pot}=144$~Ry for
  the potential. 76 $k$-points were used for the integration in the irreducible part of the Brillouin zone (BZ). To model the effect of substrate- induced strain, the in-plane lattice parameter of the superlattice was set to that of LAO ($a=3.78$\AA), STO ($a=3.905$\AA) or DyScO$_3$ (DSO) ($a=3.94$\AA), respectively. For the out-of-plane parameter we have used values obtained from XRD measurements: 3.83 (STO) and 3.93 \AA\ (LAO).~\cite{freelandcm} Relaxations of the $c$ parameter for superlattices strained at $a_{DSO}$ rendered  a $c$ value very similar to the one for the STO case. This implies a deviation from the volume conserving scenario for tensile strain, as also suggested from XRD measurements.~\cite{freelandcm} While previous DFT studies~\cite{hansmann09,millis} have used a tetragonal setup, we explored the influence of strain on the octahedral rotations by optimizing the internal parameters in a monoclinic unit cell containing 20 atoms for LNO bulk and the 1/1 superlattice. 
In the present study parallel alignment of the  Ni spins is considered. Calculations for antiferromagnetic configurations were   unstable and higher in energy. 

\section{Bulk LNO}

\begin{figure}[ht]\vspace{-0pt}
\begin{tabular}{c}
	\includegraphics[width=0.48\textwidth]{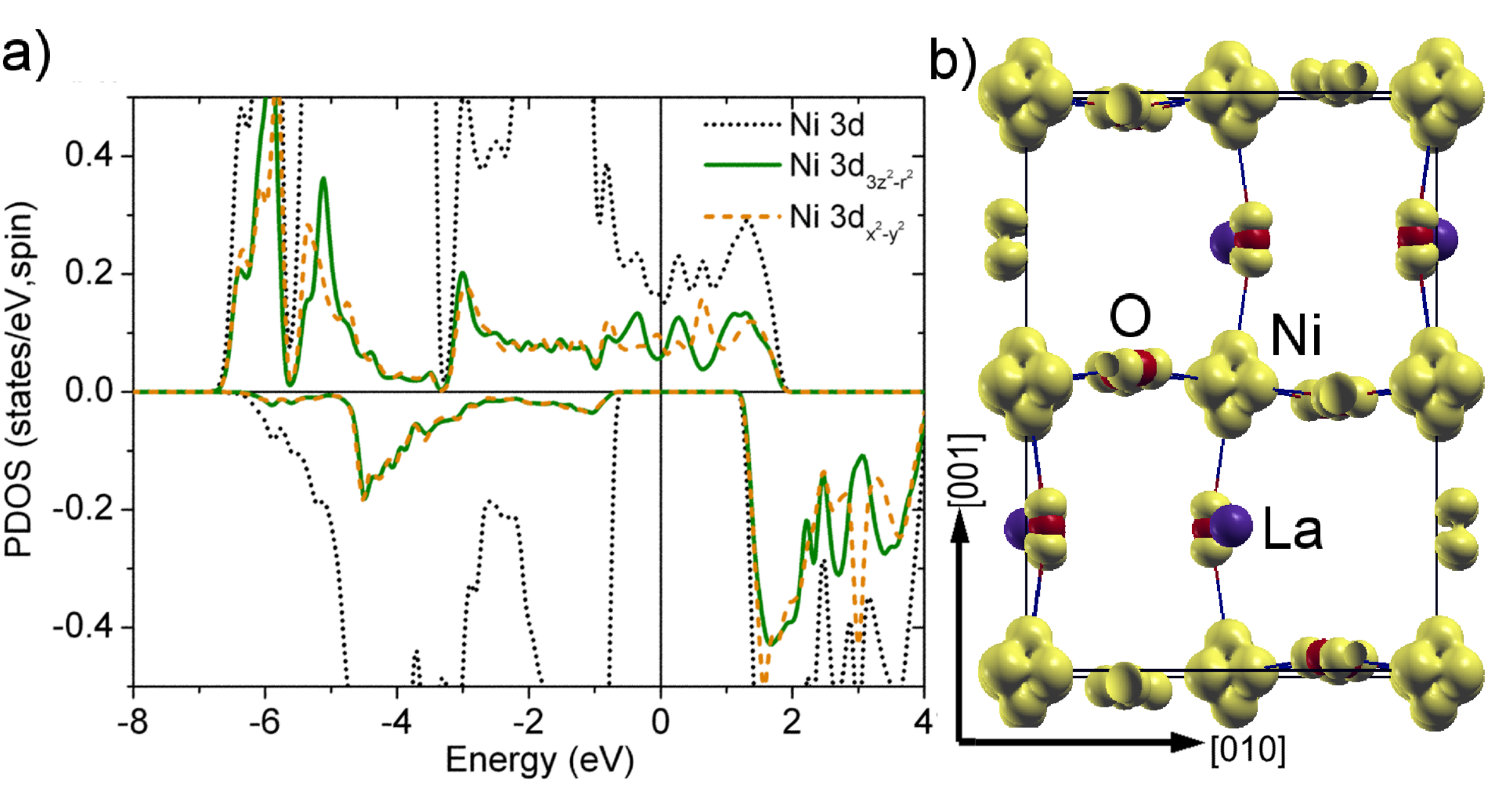}
\end{tabular}
\caption{\label{doscdnbulk}(Color online) Bulk \lno. (a) Projected density of states of Ni $3d$ character (black dotted line) obtained within GGA+$U$ ($U=4$~eV and $J=0.7$~eV). Positive (negative) values correspond to majority (minority) spin. Note that \dzt\ (green line) and \dxtyt\ orbitals (orange dashed line) are degenerate in bulk LNO. (b) Electron density distribution of the unoccupied states integrated between E$_{\rm F}$ and E$_{\rm F}$+2.5 eV. La and O ions are denoted by purple and red spheres, respectively.}
\end{figure}
Prior to studying the LAO/LNO superlattices we investigate here the electronic properties of LNO bulk. In order to facilitate comparison to the LNO/LAO superlattices the calculations are performed in a monoclinic unit cell containing 20 atoms. Figure~\ref{doscdnbulk}(a) shows the projected density of states at the Ni sites. In bulk LNO \dzt\ and \dxtyt\ are degenerate and the broad $e_g$ band  extends from -7 eV to 2 eV in the majority spin channel, while the minority $e_g$ states are not occupied except for some hybridization with the O $2p$ states between -4.5 and -3.5 eV.  The formal Ni valence in LNO is considered to be Ni$^{3+}$ ($t_{2g}^6$,$e_g^1$), but the occupation of the Ni $e_g$ states is much higher (both the \dzt\ and \dxtyt\ orbtails contain one electron each; see Table~\ref{tab:occc}).  As displayed in the  hole density distribution integrated between $E_{\rm F}$ and $E_{\rm F}+2.5$~eV [Fig~\ref{doscdnbulk}(b)], the higher number of electrons at the Ni sites is counterbalanced by holes on the ligands of $p_{\sigma}$ character ($\sim 0.2 e$ within the MT). This corresponds to a state closer to  3$d^8$L, where L denotes a hole on the oxygen site, as observed also for other rare-earth nickelates.~\cite{mizokawa95} For this reason bulk \lno\ was ruled out as a possible 
parent compound for high-Tc superconductivity.~\cite{Anisimov99}
\begin{figure}[ht]\vspace{-0pt}
\subfloat[Majority spin channel]{\label{lnomcup}\includegraphics[width=0.25\textwidth]{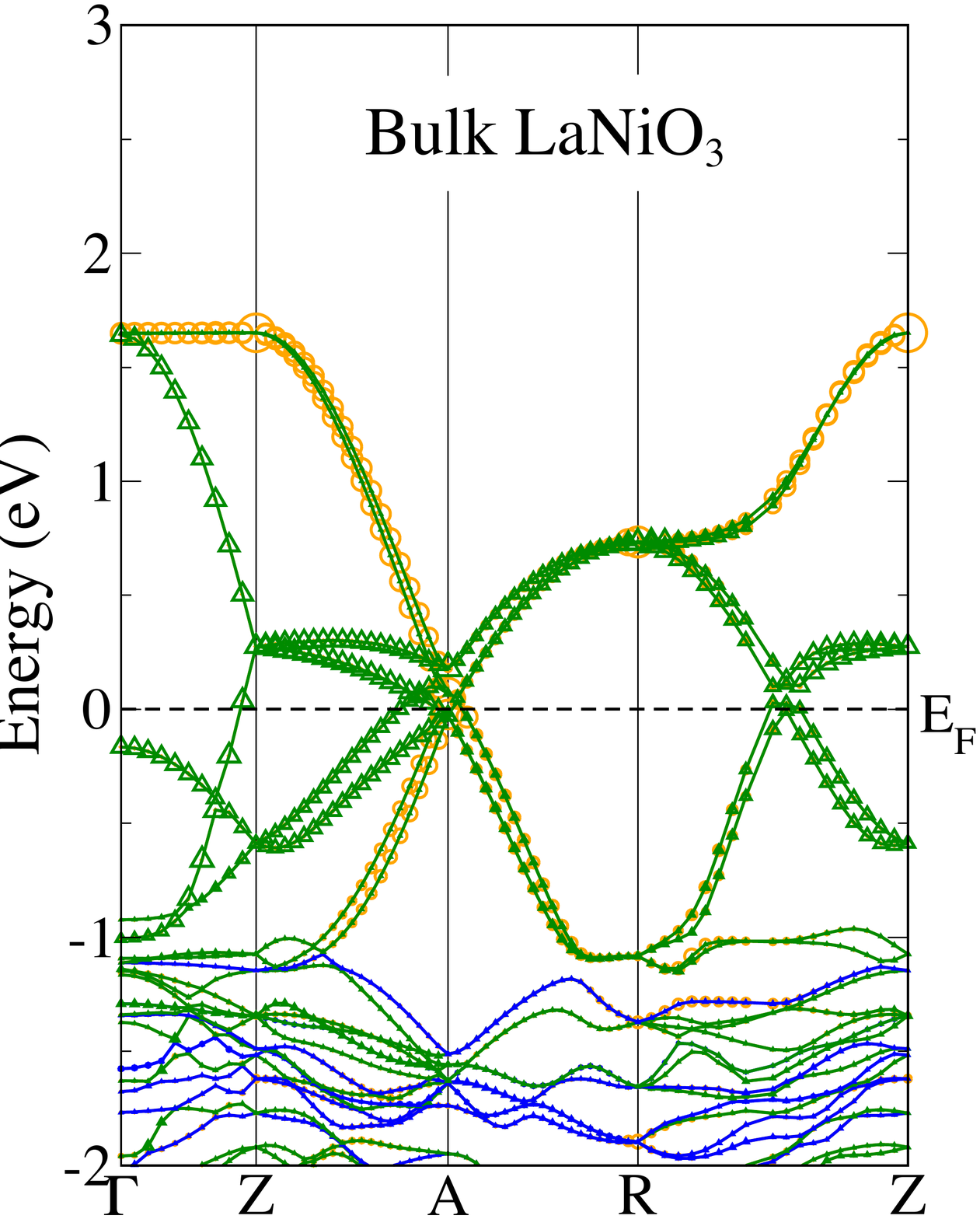}}
\subfloat[Minority spin channel]{\label{lnomcdn}\includegraphics[width=0.25\textwidth]{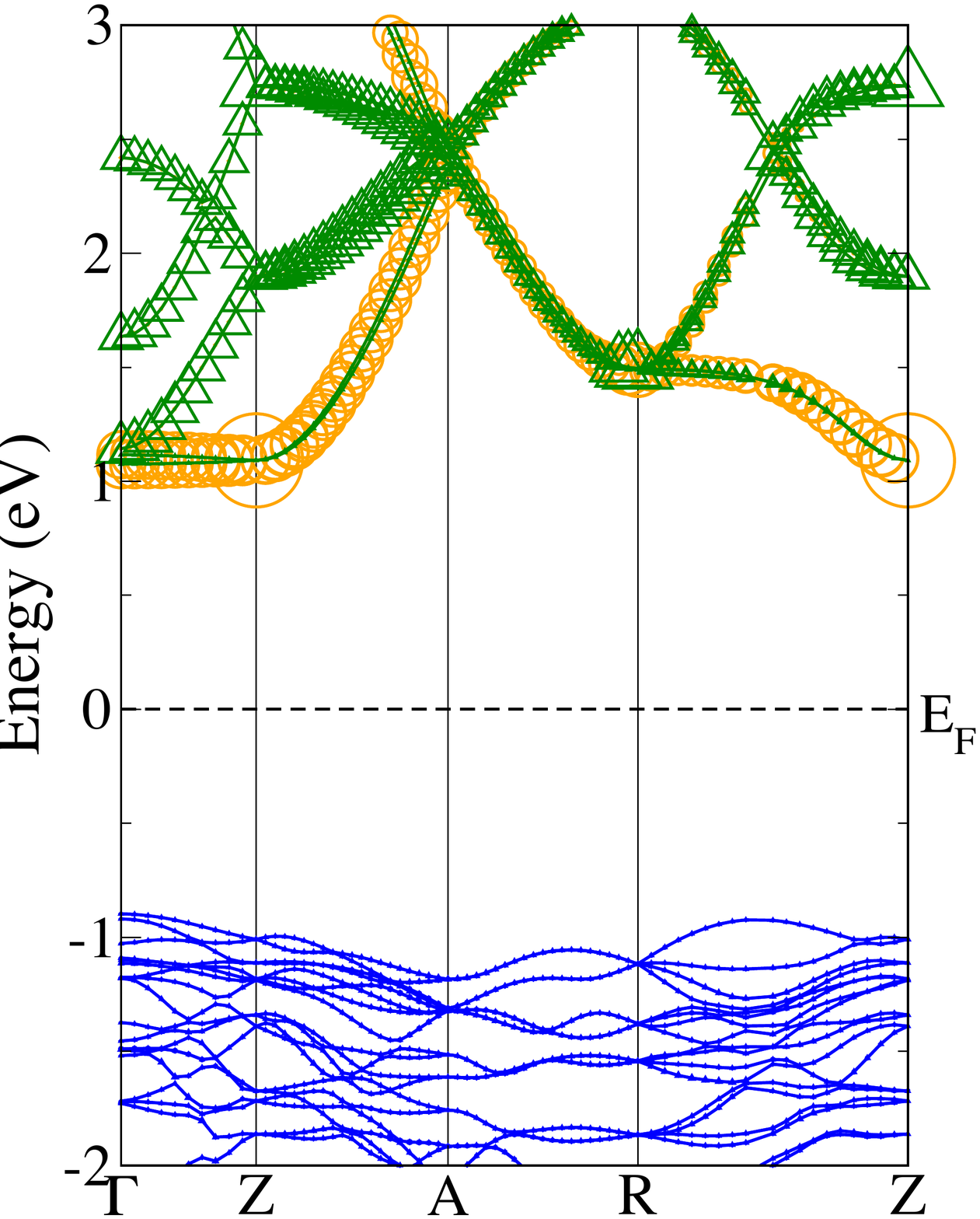}} \\
\caption{\label{fig:bulkband} (Color online) Band structure obtained within GGA+$U$ of bulk LaNiO$_3$ along the high symmetry lines $\Gamma (0,0,0)-Z (0,0,\pi)-A(\pi,0,\pi)-R(\pi,\pi,\pi)-Z(0,0,\pi)$ using a $\sqrt{2}a\times \sqrt{2}a\times2a$ unit cell with $a$ being the lattice parameter of the pseudocubic cell. In (a) and (b) the \dzt\ and \dxtyt\ character is emphasized by  green triangles and light orange circles, respectively.}
\end{figure}

The band structure of bulk LNO in Fig.~\ref{fig:bulkband} is plotted along the high-symmetry lines in the BZ of the  20-atom cell. We concentrate on the antibonding $e_g$ bands in the majority spin channel extending from -1~eV to 2 eV (in the minority spin channel these states are empty).  At $\Gamma$ both $e_g$ bands are degenerate, \dzt\ bands show a strong dispersion along $\Gamma-Z$, while \dxtyt\ bands are dispersive mainly along $Z-A-R-Z$. Both bands cross \ef\ at $A$ and \dzt\ also halfway along $R-Z$. 
\section{LAO/LNO Superlattices: Electronic properties}

\subsection{\one\ superlattices}
\begin{figure}[ht]\vspace{-0pt}
\begin{tabular}{c}
	\includegraphics[width=0.43\textwidth]{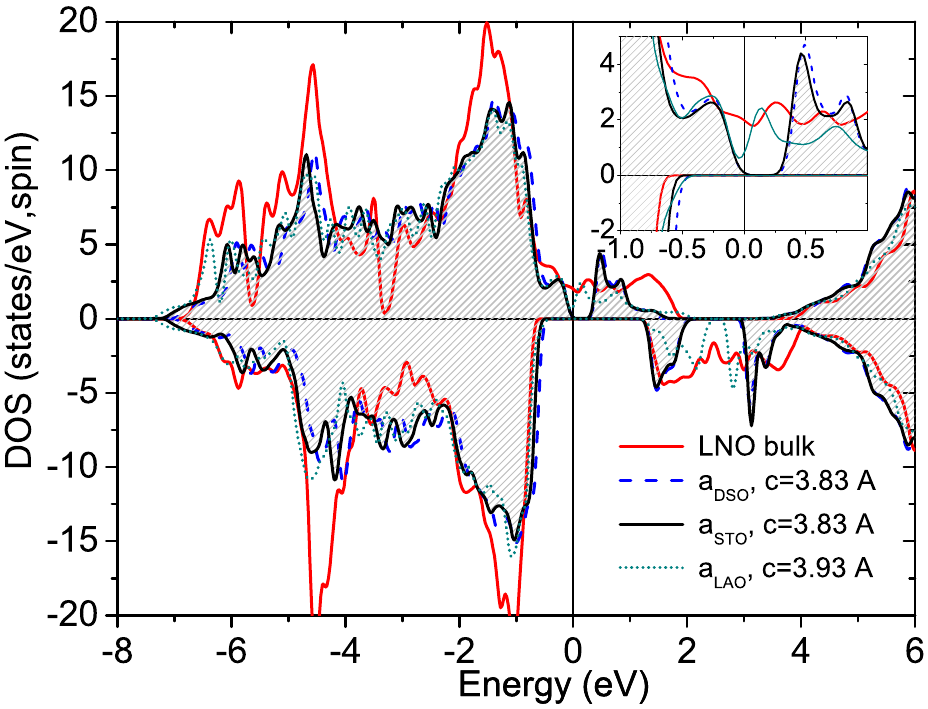}
\end{tabular}
\caption{\label{dost}(Color online) Total density of states for (LAO)$_1$/(LNO)$_1$ for the case of tensile (STO) (solid line, gray area), (DSO) (long-dashed blue line) and compressive strain (LAO) (short-dashed cyan line). Additionally, the DOS of monoclinic bulk LNO is shown with a red line.  The inset shows a zoom-in around the Fermi level  displaying the opening of a band gap for (LAO)$_1$/(LNO)$_1$ under tensile strain. }
\end{figure}
We now turn to the \one\ superlattices strained at the lateral lattice parameters of LAO, STO, and DSO. Their total density of states (DOS)  are plotted in Fig.~\ref{dost} together with the one for bulk LNO. For compressive strain (short-dashed cyan line)  the system is (semi-)metallic with a dip of  DOS at the Fermi level.  Moreover, due to the stacking of insulating LAO and conducting LNO layers an enhanced resistivity is expected for \one\ at \alao. In contrast, for tensile strain a gap in the total density of states (Fig.~\ref{dost}) of 0.3 eV occurs above the Fermi level (solid line, gaey area).  Interestnigly, the total DOS of \one\ at \adso\ almost coincides with the one  for \asto\  around the Fermi level, implying a nonlinear dependence of the electronic properties on strain. Therefore, we concentrate in the following on the electronic properties of  \one\ strained at \alao\ and \asto.

\begin{figure}[ht]\vspace{-0pt}
\centering
	 \includegraphics[width=0.2433\textwidth]{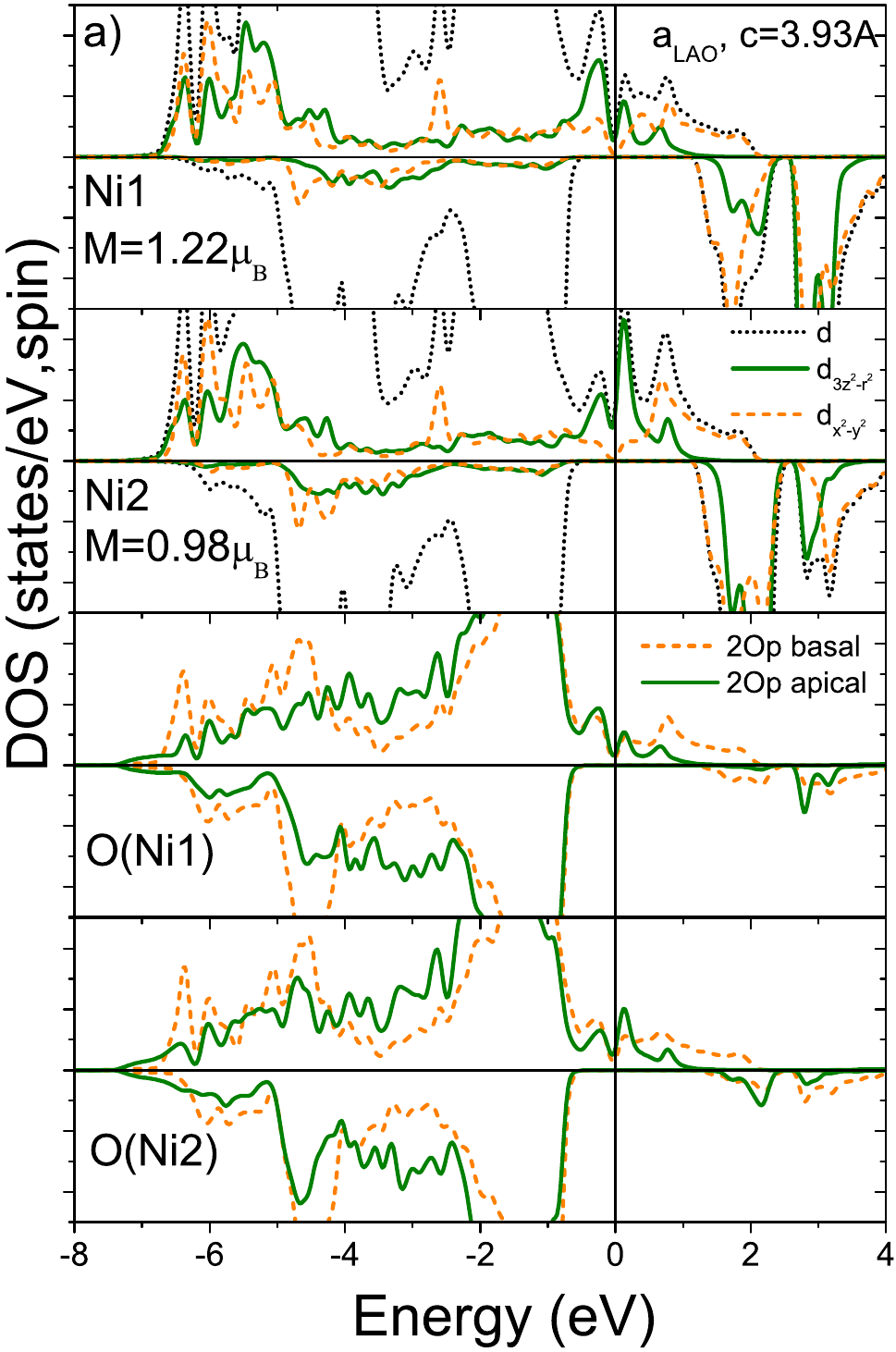}
	 \includegraphics[width=0.23\textwidth]{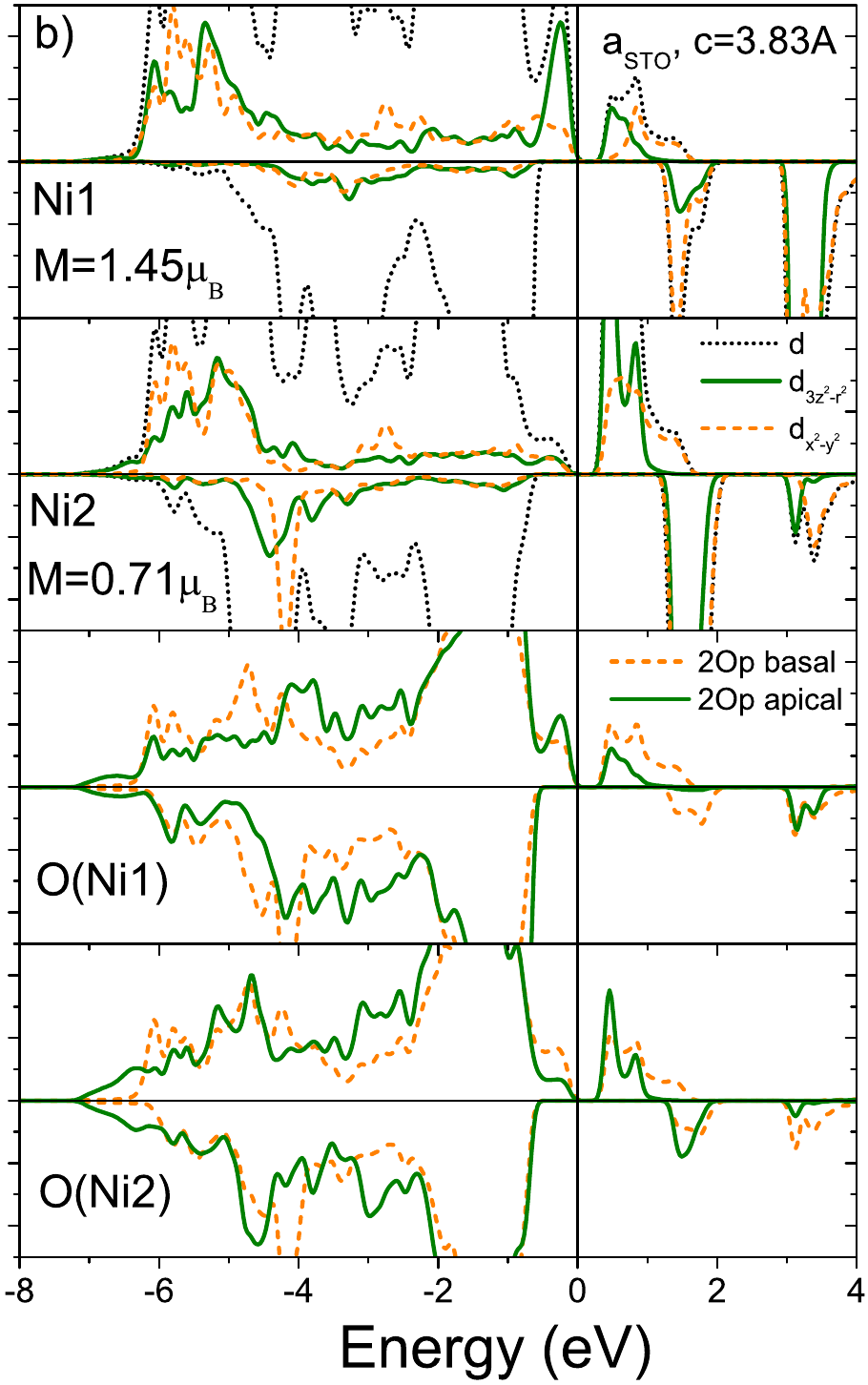}
\caption{\label{pdos11}(Color online) Projected density of states (PDOS)   of Ni $3d$ states (black dotted line), \dzt\ (green line) and \dxtyt\ orbitals (orange dashed line)  in (LAO)$_1$/(LNO)$_1$ for (a) compressive (LAO)  and (b) tensile (STO) strain. The lower panels show the PDOS of basal (orange dashed line) and apical (green line) oxygens. Note the charge disproportionation at the Ni sites and the reduced density at the apical oxygen sites. }
\end{figure}
\begin{table*}
\caption{\label{tab:occc} Orbital occupation (total and majority/minority spin) of Ni $e_g$ states in (LAO)$_n$/(LNO)$_n$ strained at $a_{\rm STO}$ (tensile), $a_{\rm LAO}$ (compressive), and $a_{\rm DSO}$ (tensile) compared to Ni in bulk LNO. The lateral strain with respect to the bulk LNO lattice parameter $a_0$ is $\epsilon=(a-a_0)/a_0$. Additionally, the orbital polarization of the occupied $e_g$ states $P = (n_{x^2-y^2}-n_{3z^2-r^2})/(n_{x^2-y^2} + n_{3z^2-r^2})$  and the  magnetic moments are displayed. The last two columns give the number of  holes $\Delta p_{\sigma}$ (difference in occupation with respect to the fully occupied $p_{\pi}$ orbitals) of the apical (a) and basal (b)  oxygen within the MT sphere.}
\smallskip
\begin{ruledtabular}
\begin{tabular}{lllrrrrr}
&& $n_{3z^2-r^2}$ & $n_{x^2-y^2}$ & P[$\%$] & M [$\mu_{\rm B}$]& $\Delta p_{\sigma}^a$ & $\Delta p_{\sigma}^b$ \\
\hline
1/1@$a_{\rm LAO}$   & Ni1 & 1.06 (0.86/0.19) & 0.98 (0.76/0.22) & -3.8&   1.22 &  -0.10 & -0.19 \\
$\epsilon=$-1.1$\%$  & Ni2 & 1.02 (0.77/0.24) & 0.98 (0.72/0.26) & -1.6&  0.98 &  -0.13 & -0.19 \\
1/1@$a_{\rm STO}$   & Ni1 & 1.05 (0.87/0.18) & 1.02 (0.86/0.15) & -1.7& 1.45 &  -0.09 & -0.22 \\
$\epsilon=$1.7$\%$   & Ni2 & 0.97 (0.63/0.33) & 1.01 (0.71/0.30) &  2.2&  0.71 &  -0.15 & -0.22 \\
1/1@$a_{\rm DSO}$   & Ni1 & 1.05 (0.87/0.18) & 1.02 (0.87/0.14) & -1.7&   1.47 &  -0.09 & -0.22 \\
$\epsilon=$2.7$\%$   & Ni2 & 0.97 (0.63/0.33) & 1.01 (0.71/0.29) &  1.9&  0.71 &  -0.16 & -0.22 \\
3/3@$a_{\rm STO}$   & Ni IF   & 1.00 (0.77/0.23) & 1.02 (0.80/0.21) & 0.8&  1.14 &  -0.12 & -0.20 \\
$\epsilon=$1.7$\%$   & Ni IF-1 & 1.02 (0.81/0.21) & 1.01 (0.80/0.21) &  -0.3&  1.20 &  -0.21 & -0.20 \\
LaNiO$_3$        & Ni  & 1.00 (0.78/0.22) & 1.00 (0.77/0.22) & 0.0&   1.12 &  -0.19 & -0.20 \\
\end{tabular}
\par\vspace{-0.05\skip\footins}
\renewcommand{\footnoterule}{}
\end{ruledtabular}
\end{table*}
In order to understand the origin of the  distinct electronic behavior for tensile and compressive strain we analyze  here the PDOS of the Ni $3d$ orbitals. Besides the filled $t_{2g}$ orbitals there is a significant occupation of both $e_g$ orbitals at the Ni sites (see Fig.\ \ref{pdos11}) split into bonding (between -7 and -5 eV)  and antibonding contributions (from -1 to 2 eV). 
The most striking effect for tensile strain is a charge disproportionation on the Ni sublattice, Ni$^{3+}\rightarrow $Ni$^{3+\delta}$+Ni$^{3-\delta}$~\footnote{Although due to covalency the Ni $3d$ band occupation  is a mixture of 3$d^7$ and 3$d^8$L ground state, we use here for clarity the formal oxidation state of Ni.}
where Ni1 has a stronger occupation of the $e_g$ orbitals. The effect is most pronounced for the majority spin states with $0.87e$/$0.86e$  (Ni1) and $0.63e$/$0.71e$  (Ni2) in the \dzt/\dxtyt\ orbitals, respectively. The reduced occupation in the majority spin channel at Ni2 is partially compensated by a stronger filling of minority states at -4.3 eV.  Altogether, the total difference in $e_g$ occupation at the Ni1 (Ni$^{3-\delta}$) and Ni2 (Ni$^{3+\delta}$) sites is $\sim$$0.2e$, but there is a significant difference in the magnetic moments 1.45\muB\ (Ni1) and 0.71\muB\ (Ni2) (for comparison, the magnetic moment in LNO bulk is 1.12\muB). Without the octahedral tilts and rotations such a state is reported at $U\geq5$~eV.~\cite{andersen} A similar tendency for disproportionation was recently found in LNO films under tensile biaxial strain,~\cite{smay,jclno} but here as a result of confinement in the LAO/LNO superlattice the effect is strongly enhanced leading to the  opening of a gap of $\sim$0.5 eV at the Fermi level.   Concerning the orbital polarization, the two sites show opposite, largely compensating tendencies: a slightly higher occupation of \dxtyt\ at Ni2 (+2.2\%)  and the reverse effect at Ni1 (-1.7\%). 

For compressive strain  the charge disproportionation is noticeably  reduced, reflected in a much smaller difference in the magnetic moments (1.22 \muB\ at Ni1 and 0.98 \muB\ at Ni2). We note that a non-disproportionated configuration lies only 12 meV/20-atom unit cell higher in energy. The broadening of the conduction band leads to an overlap with the valence band which closes the band gap with a dip in the PDOS at \ef. As a consequence of the distortion (elongation) of the NiO$_6$ octahedra discussed below, there is a stronger splitting between the \dzt\ and \dxtyt\ states around $E_F$ and an enhanced occupation of the \dzt\ orbital in the majority spin channel [$0.86e$ (\dzt) vs $0.76e$ (\dxtyt) at Ni1 and $0.77e$ (\dzt) vs $0.72e$ (\dxtyt) at Ni2]. Taking into account the occupation of the minority spin chanel the total number of electrons within the MT sphere is 1.06 (\dzt) vs. 0.98 (\dxtyt) at Ni1 and 1.02 (\dzt) vs 0.98 (\dxtyt) at Ni2, respectively. This results in an orbital polarization of $-3.8\%$ and $-1.6\%$. We note that much higher values were obtained by Han emph{et al.}~\cite{millis} by integrating only over the antibonding states from $\sim-1.5$~eV to \ef. 


\begin{figure}[ht!]
\centering
\subfloat[Majority spin channel Ni 1]{\label{laoni1up}\includegraphics[width=0.25\textwidth]{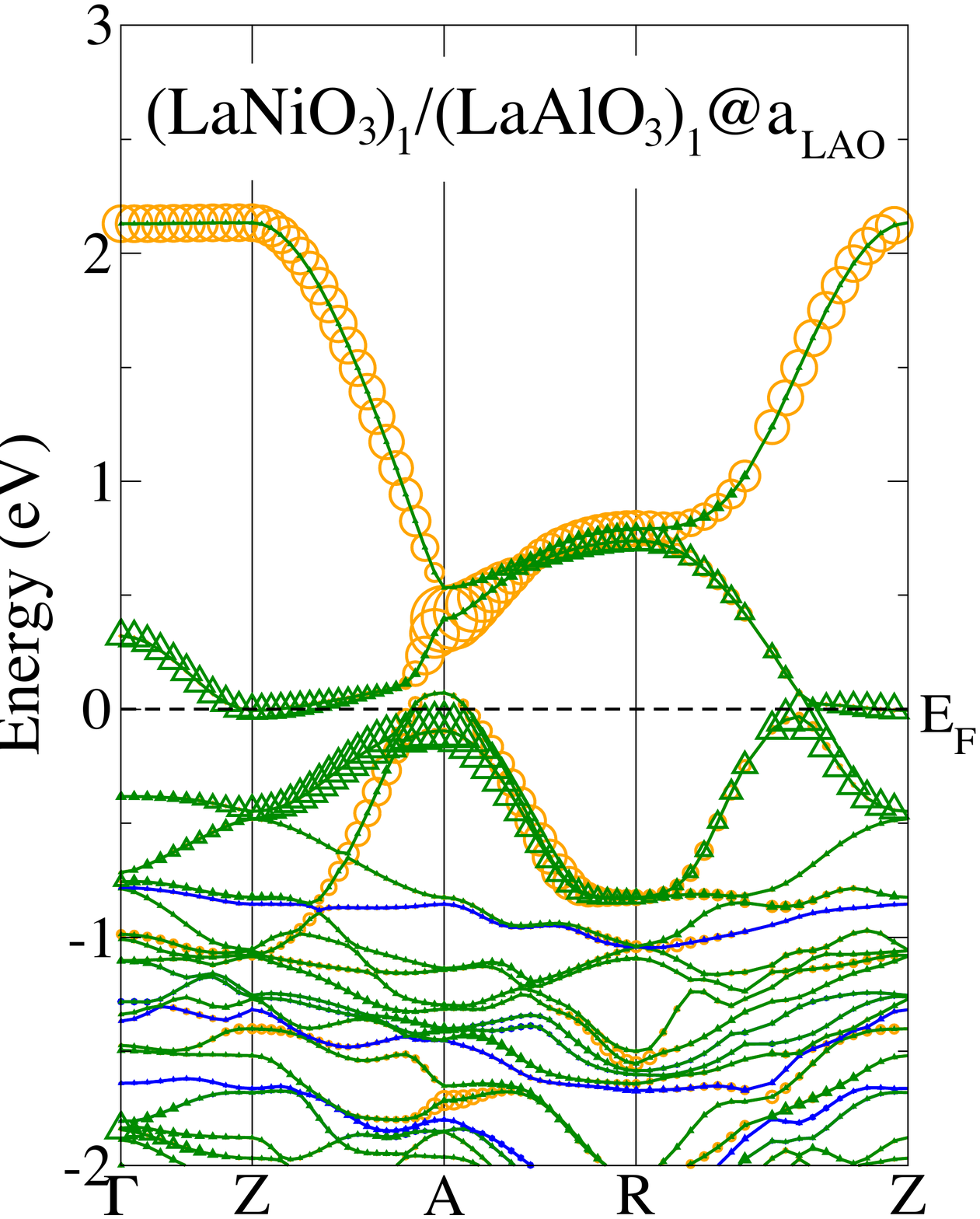}}
\subfloat[Majority spin channel Ni 2]{\label{laoni2up}\includegraphics[width=0.25\textwidth]{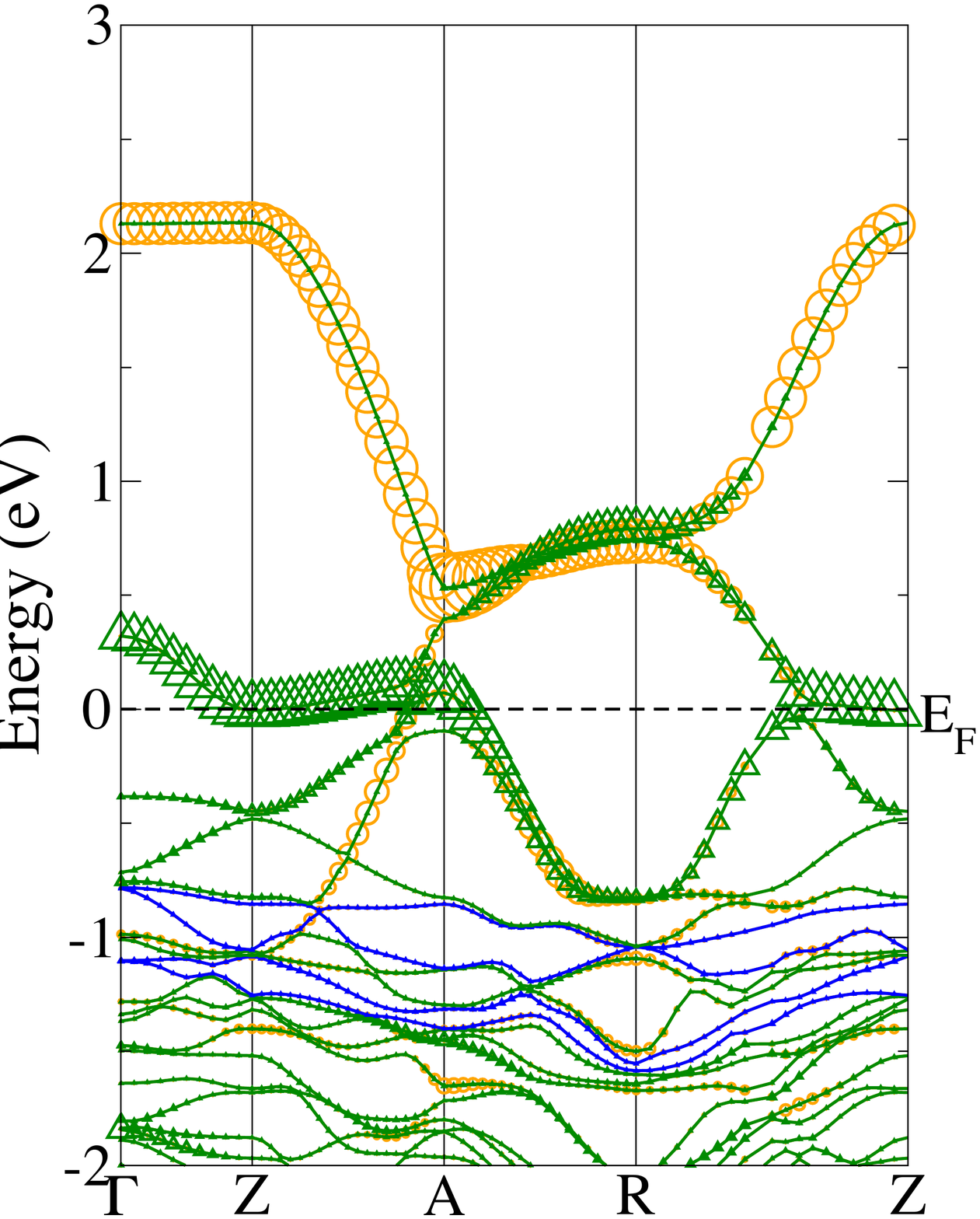}} \\
\caption{(Color online) Majority spin band structure of (LAO)$_1$/(LNO)$_1$ under compressive strain (\alao).
Note the reduced dispersion of the \dzt\ band (green triangles) compared to bulk LNO caused by confinement and the opposite effect for the \dxtyt\ band (orange circles) due to the reduction of the in-plane lattice parameter.  }
\label{fig:laoband}
\end{figure}



\begin{figure}[ht!]
\centering
\subfloat[Majority spin channel Ni 1]{\label{stoni1up}\includegraphics[width=0.25\textwidth]{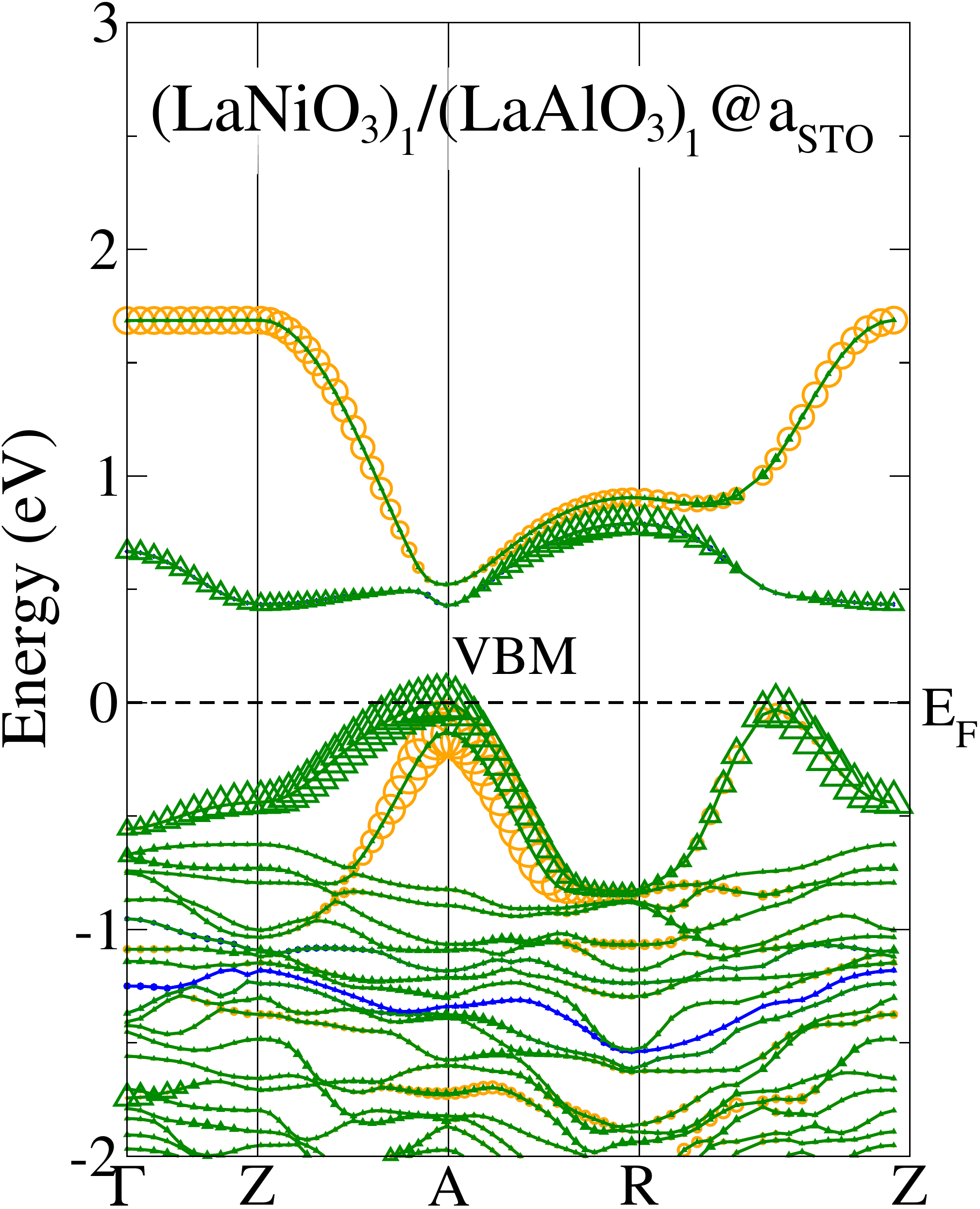}}
\subfloat[Majority spin channel Ni 2]{\label{stoni2up}\includegraphics[width=0.25\textwidth]{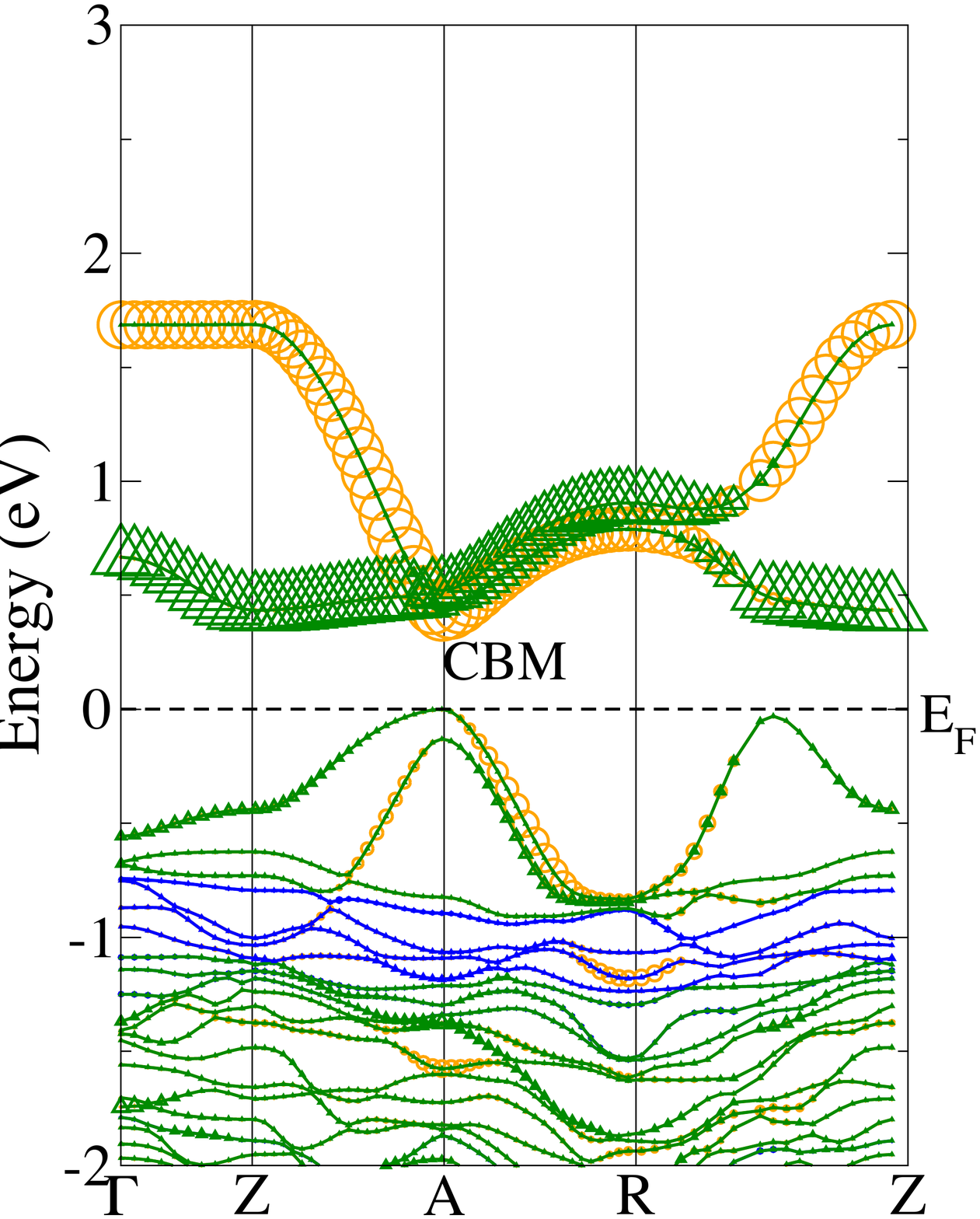}}  \\
\caption{(Color online) Majority spin band structure of (LAO)$_1$/(LNO)$_1$ under tensile (STO) strain. 
\dzt\ and \dxtyt\ character is emphasized by  green triangles and orange circles, respectively. The direct band gap opens at A with a valence band maximum (VBM) of predominantly Ni1 character and  conduction band minimum (CBM) with mostly Ni2 contribution.}
\label{fig:stoband}
\end{figure}

To gain more insight in the electronic structure of the \one\ superlattice, we have plotted in Fig.~\ref{fig:laoband} and Fig.~\ref{fig:stoband} the band structure in the majority spin channel for compressive (\alao) and tensile (\asto) strain, respectively. As a consequence of confinement in the superlattice the \dzt band has a reduced dispersion compared to bulk \lno\ (Fig.~\ref{fig:bulkband}) and the degeneracy of the two $e_g$ states at $\Gamma$ is lifted with the \dzt state being 0.3 (\alao)/0.7 (\asto) above \ef. The band structure around \ef\ is dominated by more localized \dzt\ bands and strongly dispersive \dxtyt\ bands (especially for compressive strain) with dispersion of the latter mainly along $Z-A-R-Z$. As already mentioned in the discussion of the PDOS, there is a stronger splitting between \dzt\ and \dxtyt\ states for \alao\ compared to \asto.  In contrast to bulk LNO (Fig.~\ref{fig:bulkband}) a gap opens at $A$ both for tensile and compresive strain. However, for compressive strain (\alao) the \dzt\ band lies at the Fermi level along $Z-A$ and halfway along $R-Z$ causing a finite DOS at \ef. The upward shift of this band by $\sim0.5$~eV opens a band gap for \asto. While the valence band maximum is mainly determined by \dzt\ (and to a lesser extent \dxtyt) states of Ni1 at $A$ hybridized with $p_z$ orbitals of the apical oxygen, the conduction band minimum is dominated by \dzt\ states of Ni2, admixed with \dxtyt and $p_z$ bands of the respective apical oxygen. This results in a stronger suppression of ligand holes of the apical oxygen of Ni1.


\begin{figure}[ht]\vspace{-0pt}
\subfloat[\alao]{\label{laodens}\includegraphics[width=0.23\textwidth]{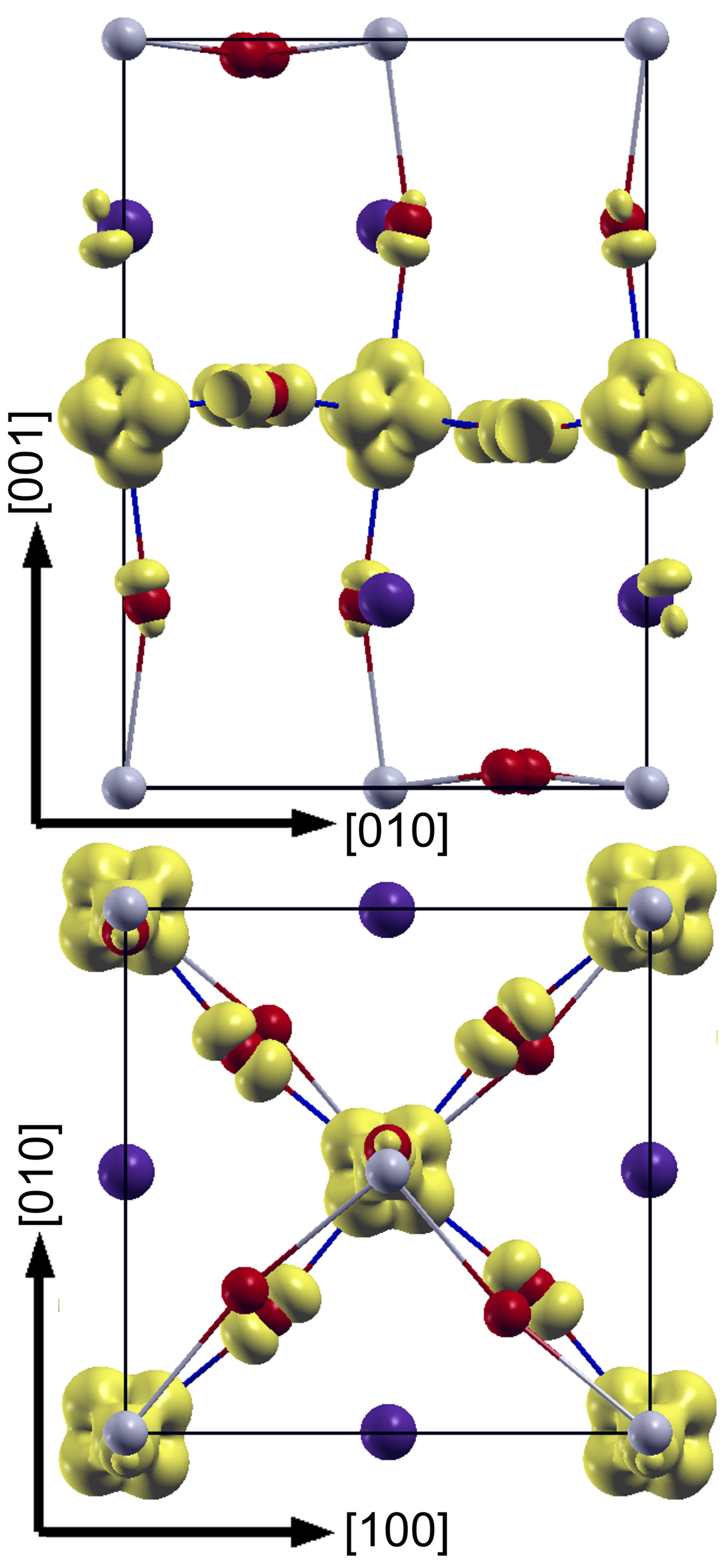}}
\subfloat[\asto]{\label{stodens}\includegraphics[width=0.226\textwidth]{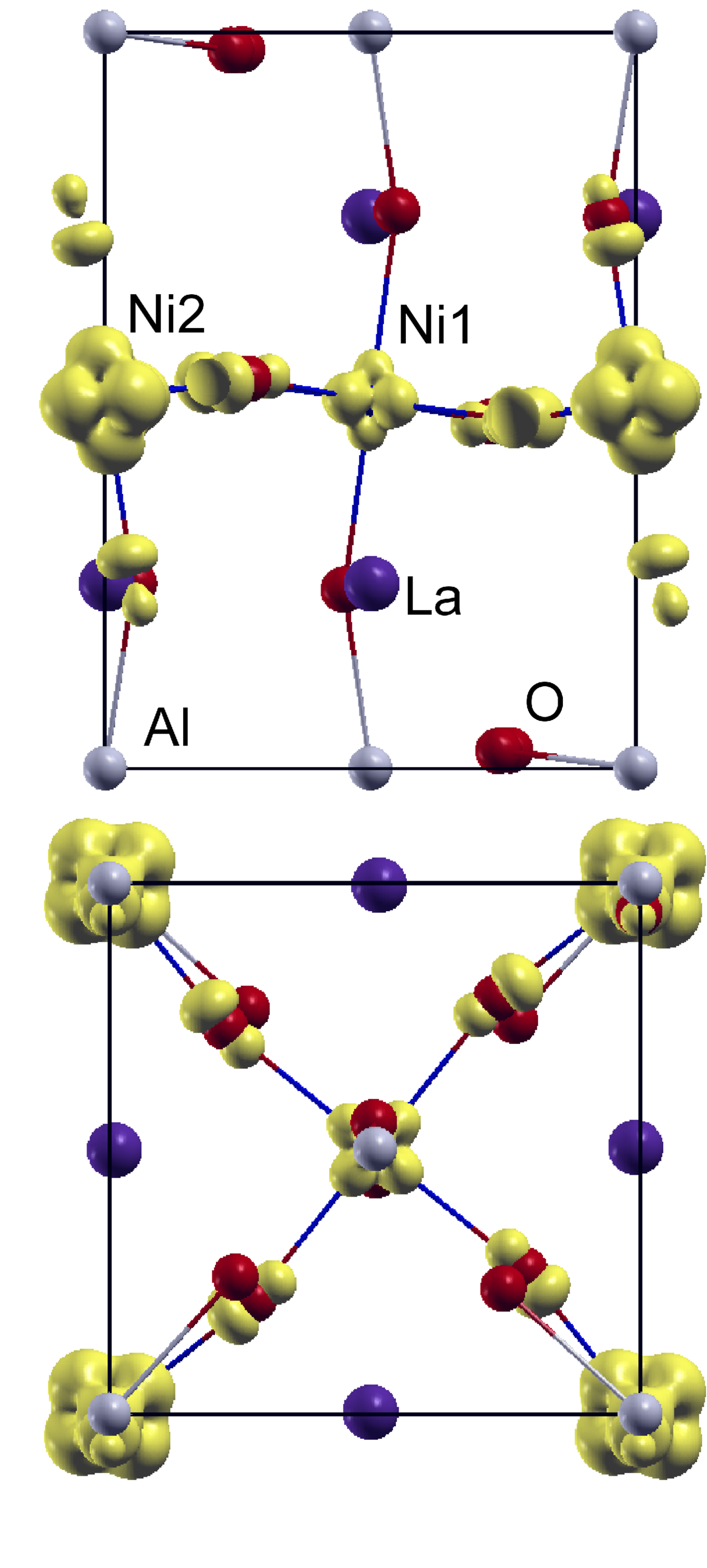}}
\caption{(Color online) Electron density distribution of the unoccupied states integrated between E$_{\rm F}$ and E$_{\rm F}$+2.5 eV in (LAO)$_1$/(LNO)$_1$ for  (a) compressive (LAO) and (b) tensile (STO) strain. Note the  reduction in hole density on the apical oxygen between Ni and Al, as well as the charge disproportionation at the Ni sites for tensile strain. La, O, and Al ions are shown in purple, red, and light gray, respectively.}
\label{11cdn}
\end{figure}

The charge disproportionation at the Ni sites for  \asto\  is visible in the spatial distribution of holes  shown in Fig.~\ref{11cdn}. In contrast to the three-dimensional network of holes on the basal and apical oxygens in bulk LNO [Fig.~\ref{doscdnbulk}(b)], a striking feature in \one\ is the two- dimensional character of hole distribution within the LNO part of the superlattice due to a reduced number of holes  in the $p_z$ orbital of the apical oxygen for both tensile and compressive strain (see also PDOS of oxygen in Fig.~\ref{pdos11}). At the same time, the in-plane oxygen orbital occupancies are very similar to those in bulk LNO. As a measure of the number of holes in Table~\ref{tab:occc} we have determined the difference in filling between $p_{\sigma}$ and the fully occupied $p_{\pi}$ orbitals. We note that we consider here only the occupation within the MT sphere, which contains   60-70\% of the total charge of oxygen.  In bulk LNO there are $0.19$ less electrons in the $p_\sigma$ than in the $p_\pi$ orbital within the MT. This number is reduced to $0.09-0.15e$ for the apical oxygens in the superlattice. The suppresion is strongest for the apical oxygen of Ni1 (Ni$^{3-\delta}$) which simultaneously has a higher \dzt\ occupation. Thus the suppression of holes due to the stronger ionicity of the Al$^{3+}$-O-bond across the interface and the charge disporportionation at the Ni sites with more ionic and more covalent contributions at the two distinct Ni-sites are the crucial mechanisms in achieving an insulating state in \one\ .


\subsection{\three\ superlattices}
\begin{figure}[ht]\vspace{-0pt}
\begin{tabular}{c}
	\includegraphics[width=0.48\textwidth]{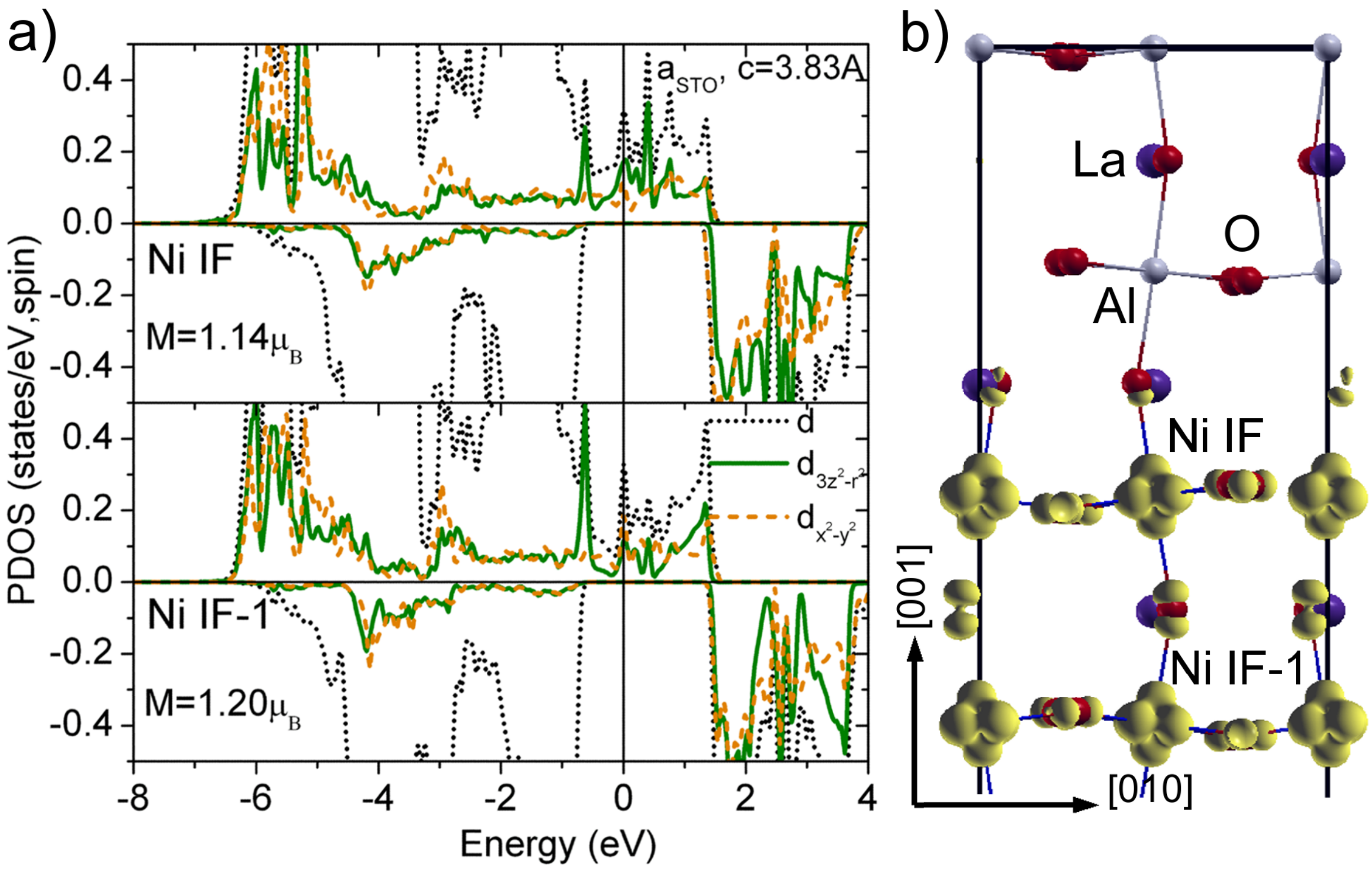}\\
\end{tabular}
\caption{\label{fig:pdoscdn33} (Color online) (a) Projected density of states (PDOS) of $3d$ states for Ni at interface (IF) and in the central layer (IF-1) in (LaAlO$_3$)$_3$/(LaNiO$_3$)$_3$ for  tensile (STO) strain. Total $3d$, \dzt\ and \dxtyt\ character are denoted by black dotted, green, and orange dashed line, respectivelyl. (b) Electron density distribution of unoccupied states (yellow) integrated between  E$_{\rm F}$ and E$_{\rm F}$+2.5 eV. Note that only half of the simulation cell is shown. For color coding see Fig.~\ref{11cdn}.}
\end{figure}

\begin{figure}[ht!]
\centering
\subfloat[Majority spin channel Ni IF]{\label{3stoniIFup}\includegraphics[width=0.25\textwidth]{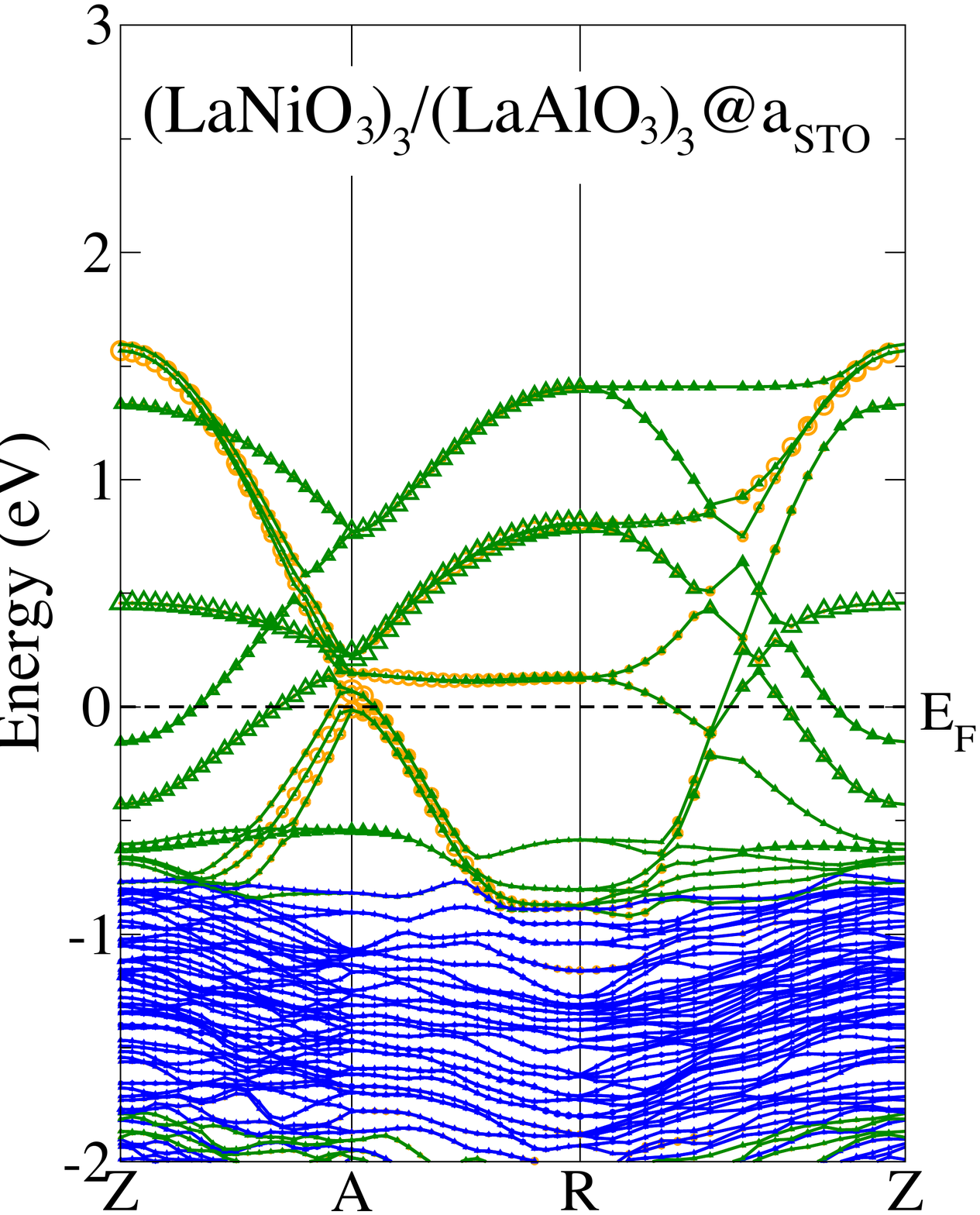}}
\subfloat[Majority spin channel Ni IF-1]{\label{3stoniIF1up}\includegraphics[width=0.25\textwidth]{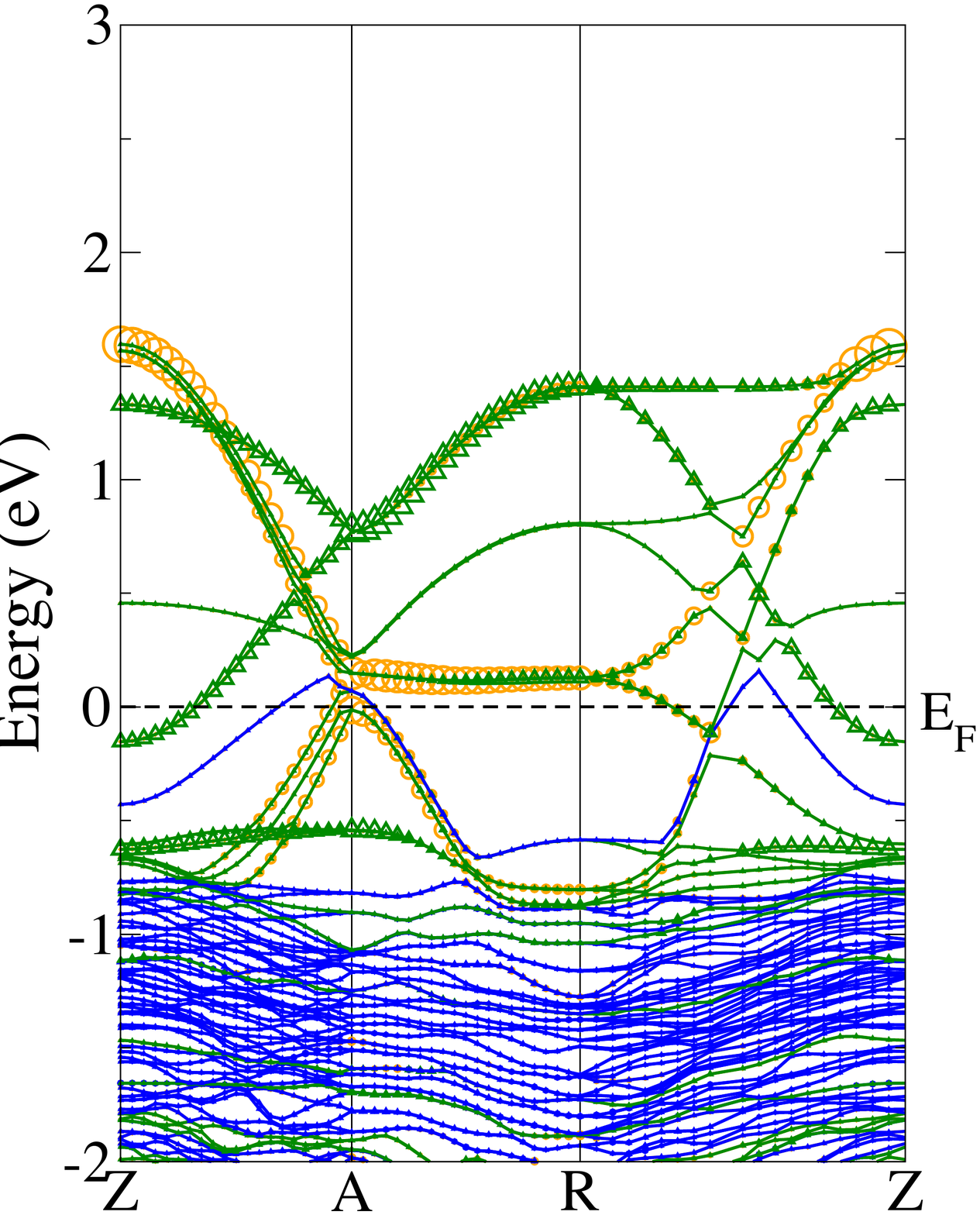}} \\
\vspace*{1ex}
\caption{(Color online) Majority spin band structure of (\lao)$_3$/(\lno)$_3$ for tensile (STO) strain. 
Note the splitting of \dzt bands (green triangles) of Ni in the interface (IF) and (IF-1) layer, respectively.  \dxtyt\ bands are emphasized by   orange circles.  }
\label{fig:band33}
\end{figure}

We turn now to the effect of LNO thickness on the electronic properties of the superlattice.
The projected density of states (PDOS) at the Ni sites in \three\ at \asto\ is displayed in
Fig.~\ref{fig:pdoscdn33}. Unlike the \one\ system,  there is a finite DOS at \ef\ at the Ni sites both in the IF and IF-1 layer and no evidence for charge disproportionation. The $e_g$ states are nearly degenerate with a similar width as in the bulk  and a small peak at \ef. Its character is predominantly \dzt\ in the interface layer (IF) and Ni \dxtyt\  in IF-1 layer. As shown in the majority spin band structure in Fig.~\ref{fig:band33}, these bands cross \ef\ at $A$ [note that in \one\ a band gap opens at $A$]. The \dzt\ bands  of the two distinct layers (IF and IF-1) are split, with the IF band being narrower and  lying lower in energy.   In contrast to recent soft x-ray reflectivity measurements on (\lao)$_4$/(\lno)$_4$~\cite{Benckiser}, we find only a small orbital polarization  both in the interface (0.8\%) and IF-1  (-0.3\%) layers. 

Figure~\ref{fig:pdoscdn33}(b) shows the hole density distribution integrated between \ef\ and
\ef+2.5~eV. Besides the similar hole density at the Ni sites independent of their position with respect to the interface, we observe again a reduced hole density at the apical oxygen at the interface. However, the O$2p_z$ hole density is quickly restored to nearly bulk LNO behavior already in the next layer. This quick recovery of bulk LNO behavior at the apical oxygens away from the interface and the metallic state at the Ni sites with no evidence for charge disproportionation are instrumental for the insulator-to-metal transition in the (LNO)$_n$/(LAO)$_n$ SLs with increasing  $n$, consistent with recent experiments.~\cite{liu2011,keimer}


\section{Strain and confinement induced lattice deformations}
\begin{table*}
\caption{\label{tab:bond} Ni-oxygen out-of-plane ($d_{Ni-O^a}$) and in-plane ($d_{Ni-O^b}$) distances, out-of-plane bond angle $\theta_c$ (Al-O-Ni),  and in-plane angle $\theta_{ab}$ (Ni-O-Ni). Additionally, the rotation $\theta$ and tilt $\phi$ angles, as well as Glazer's rotation angles $\alpha$ and $\gamma$, are displayed. }\smallskip
\begin{ruledtabular}
\begin{tabular}{lllrrrrrrr}
&& $d_{Ni-O^a}$ & $d_{Ni-O^b}$ & $\theta_c$ & $\theta_{ab}$ & $\theta$ & $\phi$ & $\alpha$ & $\gamma$\\
\hline
1/1@$a_{\rm LAO}$    & Ni1 & 2.02 & 1.93 & 166.9 & 162.5 & 6.2 & 6.5 & 4.6 & 6.2 \\
$\epsilon=$-1.1$\%$  & Ni2 & 2.00 & 1.91 & 166.4 & 162.5 & 6.2 & 6.8 & 4.8 & 6.2 \\
1/1@$a_{\rm STO}$    & Ni1 & 1.99 & 2.02 & 163.7 & 162.4 & 6.1 & 8.1 & 5.8 & 6.1 \\
$\epsilon=$1.7$\%$   & Ni2 & 1.93 & 1.93 & 162.0 & 162.1 & 6.1 & 9.0 & 6.4 & 6.1 \\
1/1@$a_{\rm DSO}$    & Ni1 & 2.00 & 2.05 & 160.7 & 161.7 & 6.2 & 9.7 & 6.8 & 6.2 \\
$\epsilon=$2.7$\%$   & Ni2 & 1.94 & 1.95 & 158.3 & 161.0 & 6.2 & 10.8 & 7.7 & 6.2 \\
3/3@$a_{\rm STO}$    & Ni IF   & 1.95 & 1.97 & 163.9 & 162.3 & 6.2 & 8.1 & 5.7 & 6.2 \\
$\epsilon=$1.7$\%$   & Ni IF-1 & 1.97 & 1.97 & 162.4 & 162.6 & 6.4 & 8.8 & 6.2 & 6.4 \\
LaNiO$_3$            & Ni  & 1.94 & 1.94 & 164.8 & 164.8 & 5.4 & 7.6 & 5.4 & 5.4 \\
LaAlO$_3$            & Al  & 1.90 & 1.90 & 171.4 & 171.4 & 3.0 & 4.3 & 3.0 & 3.0 \\
\end{tabular}
\par\vspace{-0.05\skip\footins}
\renewcommand{\footnoterule}{}
\end{ruledtabular}
\end{table*}
\begin{figure}[ht]\vspace{-0pt}
\centering
\includegraphics[width=0.4\textwidth]{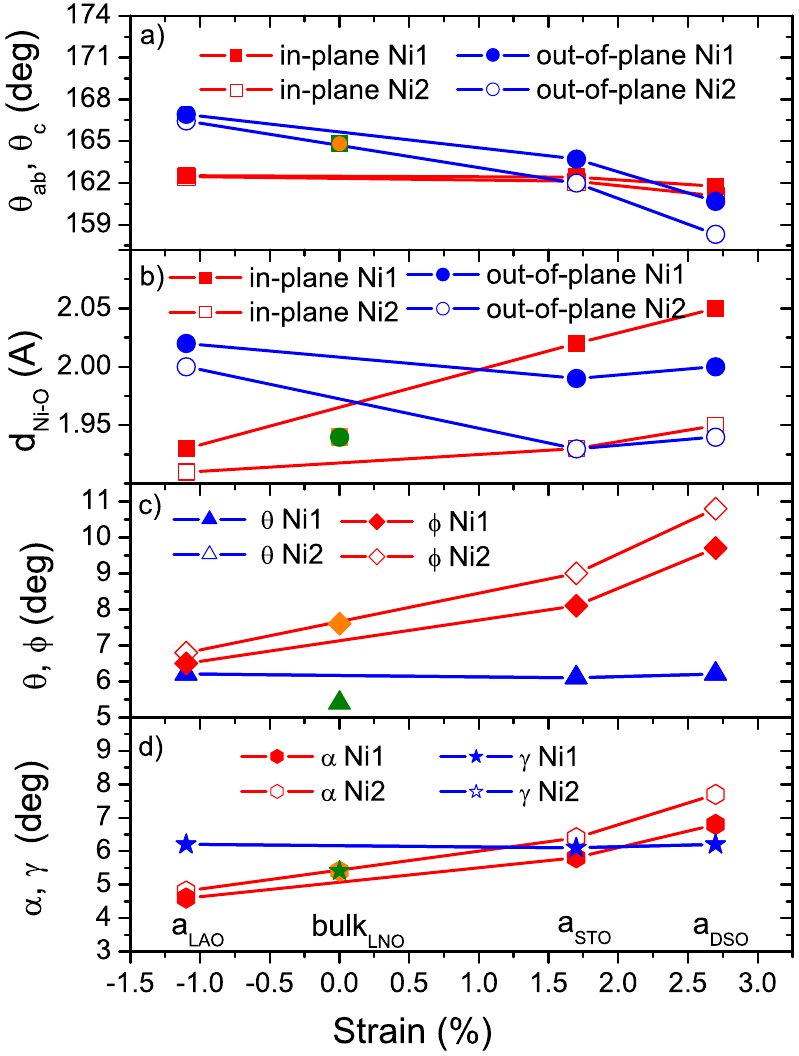}\\
\caption{\label{fig:angles}(Color online) Strain dependence of the in-plane (red squares) and out-of-plane (blue circles)  (a) bond angles and (b) Ni-O bond lengths in \one; (c) NiO$_6$ octahedral rotation ($\theta$) and tilt ($\phi$) angles, and (d) Glazer's rotation angles $\alpha$ and $\gamma$. The symbols at $0\%$ strain represent the  bulk LNO  values. 
} 
\label{angles}
\end{figure}
Not only the electronic but also the structural properties of the superlattice show a strongly  asymmetric response to strain. The  bond lengths and angles of the superlattices in comparison to bulk LNO and LAO are displayed in Table~\ref{tab:bond} and Fig.~\ref{fig:angles}.   For compressive strain the in-plane Ni-O bond length (1.92\ \AA) in \one\ is significantly shorter than the out-of-plane distance (2.01\ \AA), which is consistent with a volume-conserving tetragonal distortion and a crystal-field splitting between the \dzt\ and \dxtyt\ orbitals observed in Fig.~\ref{pdos11}(a). In contrast, the charge disproportionation for tensile strain goes hand in hand with bond disproportionation driven by a {\sl breathing} mode with two nearly regular octahedra with in-/out-of-plane cation-anion distances of 2.02/1.99\ \AA\ and  1.93/1.93\ \AA\ at the Ni1 and Ni2 sites at \asto, respectively.
  
As mentioned previously, in the bulk both LAO and  LNO attain a rhombohedral structure with rotations of the octahedra around all three pseudocubic axes by $\alpha=\beta=\gamma=3.04^{\circ}$ (LAO) and  $5.4^{\circ}$ (LNO)~\cite{garcia92}. This corresponds to Glazer's notation~\cite{glaser} ($a^-,a^-,a^-$). 
Recent DFT calculations~\cite{hatt} have found that under compressive strain the in-plane rotations of LAO are suppressed and the rotation pattern changes to ($a^0,a^0,c^-$). Vice versa for tensile strain the rotations are around the $[110]$ direction ($a^-,a^-,c^0$). In contrast,  an ($a^-,a^-,c^-$) pattern was reported in a combined XRD and DFT study~\cite{smay} on strained LNO films with a strong reduction of $\gamma=0.3\pm0.7^{\circ}$ and enhanced $\alpha=7.1\pm0.2^{\circ}$ for tensile (\asto) and an opposite trend for compressive (\alao) strain ($\alpha=1.7\pm0.2^{\circ}$ and $\gamma=7.9\pm0.9^{\circ}$). While the  general pattern of octahedral rotations is preserved in the \one\ superlattice, the presence of the interface and the different strain state in the LAO and LNO part imposed by the substrate lead to significant changes of the absolute values of rotation angles compared to the individual phases described above. The rotational angle of the NiO$_6$ octahedra around [100] increases with strain from $\alpha \sim 4.7^{\circ}$ (\alao) to $\sim 6.1^{\circ}$ (\asto) and $\sim 7.2^{\circ}$ (\adso), whereas  $\gamma$ remains constant $\sim6.2^{\circ}$. On the other hand, for the AlO$_6$ octahedra the rotational angle around [001] changes from  $\gamma=4.4^{\circ}$ (compressive) to $\gamma\sim 2.5^{\circ}$ (tensile strain).

The cation-oxygen-cation bond angles have a direct influence on the band width. In bulk LNO the Ni-O-Ni angle is 164.8$^{\circ}$. In LAO, the deviation of the Al-O-Al angle from 180$^{\circ}$ is smaller (171.4$^{\circ}$). For compressive strain: the out-of-plane Ni-O-Al angle in \one\ is increased to $\sim166-167^{\circ}$, while the in-plane Ni-O-Ni angle decreases to $163^{\circ}$ (the in-plane Al-O-Al angle, 165.6$^{\circ}$, is also reduced w.r.t. the bulk value). In contrast, for tensile strain both the in- (Ni-O-Ni) and out-of-plane (Ni-O-Al) bond angles are significantly reduced to $\sim162^{\circ}$ for \asto\ and  $\sim161-162^{\circ}$ for \adso. The Al-O-Al is 167.3$^{\circ}$, again smaller than the corresponding angle in bulk LAO (171.4$^{\circ}$). The decrease of bond angles reduces the  band width of \dzt\ and \dxtyt\ bands between \ef\ and \ef+4~eV, as can be seen in the PDOS in Fig.~\ref{pdos11}b). The Ni-O-Ni angle has been previously identified as the control parameter of the MIT in rare earth nickelates, where a smaller deviation from $180^{\circ}$ favors Ni $3d$-O$2p$ hybridization and a metallic state, while a strong deviation leads to a MIT~\cite{medarde97}. Indeed, the values we obtain for tensile strain ($160-162^{\circ}$) are close to the Ni-O-Ni angles in PrNiO$_3$ (158.7$^{\circ}$) and NdNiO$_3$ (157.1$^{\circ}$)~\cite{garcia92} that undergo a MIT. 

In the \three\ superlattice under tensile strain (\asto) the octahedra are nearly regular with an in-/out-of-plane Ni-O bond length of 1.97/1.95
\AA\ (IF) and 1.97/197 \AA\ (IF-1), respectively. The pattern of octahedral rotations is preserved with almost identical absolute values of rotation angles compared to the \one\ at \asto. The out-of-plane Ni-O-Al ($164^{\circ}$) and Ni-O-Ni ($162^{\circ}$) and in-plane Ni-O-Ni angles $\sim 162^{\circ}$ (IF)/$163^{\circ}$ (IF-1) are slightly higher than the ones for the  \one\ superlattice in line with the metallicity of the system.

\section{Summary}
While in the bulk LNO is metallic at all temperatures with degenerate $e_g$ orbitals, we find that quantum confinement together with tensile strain allows to stabilize an insulating state in a \one\ superlattice. The  charge disproportionation  at the Ni sublattice with one strongly and one weakly magnetic Ni ion is fostered by the lattice response to strain, expressed in decreasing bond angles and enhanced tilt and rotation angles that reduce the band width. The charge ordered state in the \one\ SL under tensile strain is characterized by  two nearly regular octahedra with distinct bond lengths and hence a reduced crystal field splitting at the Ni-sites.  On the other hand, under compressive strain bands broaden and close the band gap  and CD is strongly inhibited.  Expansion of the out-of-plane Ni-O bond length results in a  crystal-field splitting between the \dxtyt\ and \dzt\ orbitals and an energy lowering of the latter [see Fig.~\ref{pdos11}(b)].   This is similar to previous results for LNO films under compressive strain.~\cite{smay}

LAO/LNO superlattices and strained LNO films show similar tilt patterns but different magnitude of octahedral rotations. This gives further evidence that the lattice response to strain in superlattices is very different from the one in thin films and thus provides a further possibility to tune the band width and the electronic behavior in transition metal oxides, as recently reported for LNO/SMO SLs.~\cite{may2011} 

The large band width of the $e_g$ states precludes significant changes in the orbital occupation via strain. The lack of lifting of orbital degeneracy is similar to other nickelates, e.g., NdNiO$_3$ where the MIT is not associated with orbital ordering.~\cite{scagnoli,garcia09} Recently, Mazin emph{et al.}~\cite{mazin07} proposed that the CO state in rare-earth nickelates gets gradually suppressed under pressure due to overlap of bands. In the confined \one\ we observe an analogous ``melting" of the CD state  as a function of strain: The significant difference in magnetic moments at the two distinct sites Ni1 and Ni2 sites at \asto\ is gradually reduced for compressive strain.

The insulating state in the \one\ superlattice is achieved through a combination of charge disproportionation and  a reduction of the  hole density on the apical oxygens at the interface due to the ionic bonding to Al. The metallic behavior at the Ni sites and the hole density of the apical oxygen in inner layers  is quickly restored in \three\ superlattices,  in line with recent observations of a MIT with increasing LNO thickness.~\cite{liu2011,keimer} We note that due to the presence of a surface, the mechanism of MIT in single layer films~\cite{triscone2011,stemmer2010} may be different and requires further investigation.


R.P. acknowledges discussions with  J. Freeland,  J. Chakhalian, B. Keimer and R. J. Angel and thanks O. K. Andersen for several discussions on the band structures.  We acknowledge financial support by the DFG (TRR80) and a grant for computational time at the Leibniz Rechenzentrum.

\FloatBarrier

\end{document}